# An efficient method for performance improvement of organometal halide perovskite solar cell via external electric field

Xiu Gong[a, b], Heng Ma*[a, c], Yu-Rong Jiang[a], Meng Li[b], Zhao-Kui Wang[b], Tetsuo Soga[c]

[a] Henan Key Laboratory of Photovoltaic Materials, College of Physics and Electronic Engineering, Henan Normal University, Xinxiang, Henan 453007, China

[b] Soft Materials (FUNSOM), Soochow University, Suzhou, Jiangsu 215123, China
[c] Department of Frontier Materials, Nagoya Institute of Technology, Nagoya, Aichi 466-0061, Japan

Address all correspondence to the author. *Email: hengma@henannu.edu.cn

Abstract

An effective method, performed adding external electric field (EEF) on $CH_3NH_3PbI_{3-x}Cl_x$ (OPIC) perovskite layer during the annealing process, is proposed to improve the performance of the solar cell. By harmonizing EEF direction with the hole/electron modified layer, a significant improvement on the short circuit current and fill factor is obtained. Using the simplest planar device, the largest positive EEF of $2.5\times10^6$ V/m makes PCE increase from 12.86 to 14.33, whose increment reaches 11.4% compared with non-EFE sample. By analyzing the best and the statistics data, a fine positive correlation between EEF and PEC is found. The physical mechanism which a displacement polarization field induced by the ionic migration enhances the built in field of the perovskite heterojunction is discussed. The study proposed a physical process in modifying the cell efficiency and provides a new evidence on current-voltage hysteresis of OPIC devices.

1. Introduction

Solar cell based on perovskite organic-inorganic hybrid materials is the most striking photovoltic device in past three years, which is being widely carried out research of a new photovoltaic devices. Currently, a solid planar heterojunction device has achieved near 20% power conversion efficiency (PCE) [1-3]. A substantive progress in this research possesses a potential to have a huge impact on the whole solar energy science and technology.

Aiming some key factors, researchers have made numbers of exciting jobs in materials, film growth control, hole/electron modifying layer, the electro-optic properties of interface, and the electrode interface, etc [4-7]. Currently, a crucial problem which is called hysteresis behavior happened in current density–voltage (JV) measurement is mystifying scientists, because it causes that the measurement curve do not coincide, PCE reproducibility is bad, and the statistical variation of the performance parameters of the cells becomes large [8-10]. Researchers have speculated the phenomenon being rooted from light-absorbing layer in the following fields: the capacitor charge and discharge effects[9,11]; ionic migration[8]; the ferroelectric polarization effect[12-14]; or even the carrier trapping/de-trapping in the inter-facial surface[10]. Therefore, to solve this problem will be useful for improving the performance of the perovskite solar cell.

The perovskite material $CH_3NH_3PbI_{3-x}Cl_x$ (OPIC) has been well used in organometal halide perovskite solar cell. Perovskite material has been identified as ferroelectric material, so OPIC is naturally considered that the polarization-induced ferroelectric field should play an important role in JV hysteresis[12-14]. On the other hand, the ionic migration of the perovskite crystal cell can also result in the similar effects under a external bias[15,16]. A recent study has denied the ferroelectricity and, meanwhile offered important evidence of the ionic migration in modifying the efficiency of OPIC devices[17].

In our opining, whatever for ferroelectricity or for ionic migration, the polarization of the perovskite layer is a key factor on JV measurement. We have investigated the polarization effect of a bulk heterojunction made from poly-3(hexyl-thiophene) (P3HT) and [6, 6]-phenyl-C61 butyric acid methyl ester (PCBM) when an external electric field (EEF) was applied during the its annealing process[18]. We concluded that the displacement polarization of the active layer was induced EEF because of the polar PCBM and P3HT. For the up direction of EEF which can enhance the built in field (BIF) of the heterojunction, a significant improvement on the solar cell performance was obtained.

In annealing process, the polar perovskite crystal unit is readily movable and certainly exhibits a polarization in EEF. In this study, a novel method which different EEFs are added on the perovskite layer to induce polarization is proposed. In direction harmonized enhancing the BIF of the perovskite heterojunction, an ideal result of increasing PCE of the solar cell is obtained.

2. Experimental

2.1. Materials and Preparation.
Main Materials: Poly(3,4-ethylene-dioxythiophene) : polystyrenesulfonate (PEDOT:PSS) was purchased from Heraeus (Germany); Lead(II) chloride ($PbCl_2$) (99.999%) and anhydrous N,N-dimethylformamide (DMF) (99.9%) were purchased from Alfa-Aesar; $PC_{61}BM$ and Bphen were produced by Nichem Fine Technology Co., Ltd. (Taiwan).

Synthetic Methylamine Iodide (CH3NH3I): 20 mL of hydroiodic acid (57 wt % in $H_2O$) was drop wise added to 48mL(40 wt % in methanol) under ice bath stirring for 2 h. The reactants solution was distill in the rotary evaporator at 55 °C to remove the solvents, and then the precipitate was washed by diethyl ether for 3 times. Finally, white-colored powder was collected and dried at 60 °C for 24 h in vacuum.

Perovskite Precursor Solution Preparation: the mixture of $PbCl_2$:$CH_3NH_3I$ with 1:3 molar ratio were dissolved in DMF and then stirred at 60 °C overnight.

2.2. Device and Characteristics.

The used cell structure is shown in Figure 1. ITO-coated glass substrates (～15 Ω/sq) was ultrasonicated in acetone and ethanol at room temperature for 15 min, then UV-Ozone cleaner for 15 min. Film (～45 nm thick) of PEDOT:PSS was spin-coated onto ITO substrate at 4500 r.p.m. and annealed at 140 °C for 10 min. 30wt% OPIC was spin-coated at 4000 r.p.m.

In the annealing process of OPIC, EEF was exerted on the perovskite film using another ITO substrate cover, where a gap of 1mm or so was set. EEF was generated by a DC power source which the voltages were set by 50, 100, and 250 volt, respectively. EEF treatments were divided into two direction named up (+) and down (−), which is shown in Fig. 1. By lowering the temperature from 100 ℃ to room temperature, EEF was kept constant.

PCBM layers were deposited from a 20mg/mL chlorobenzene solution at 2000 r.p.m. Then 0.5 mg/mL Bphen in absolute ethanol was coated onto PCBM layer at 4000 r.p.m. Finally, 100 nm thick Ag (mask area of 7.25 $mm^2$) was deposited on top of the Bphen layer by thermal evaporation under $10^{-7}$Torr.

J–V characteristics of PSCs were recorded under 1 sun illumination using a programmable Keithley 2400 source meter under AM 1.5G simulated solar light.

All the process of the cell manufacturing can be found in our previous publications[19,20].

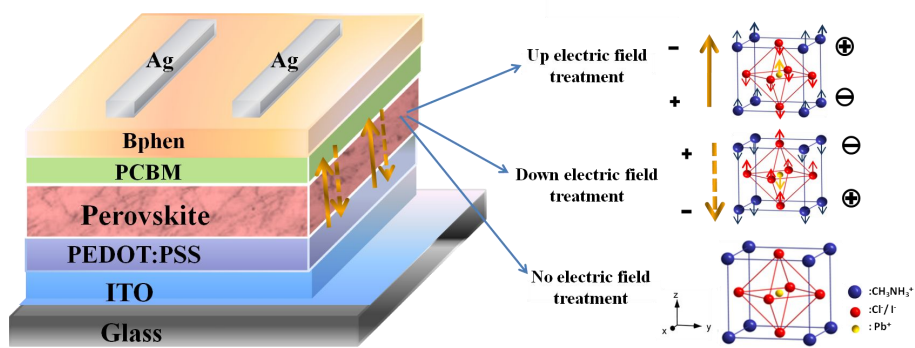

Figure 1 Cell structure and polarization conception diagram

3 Results and discussion

For perovskite solar cells, the stability and JV behavior are difficult assurance, so the same manufacturing conditions can produce different cell parameters. Generally, the cell PCE is the comprehensive assessment index. Therefore, the parameters of the cells who hold the best PCE and the average parameters in every group were collected in Table I. JV measurement and the correlations between parameters and EEFs are illustrated in Figs. 2-4, respectively.

Table 1 Photovoltaic parameters of perovskite devices with different EEFs

| External electric field (V/m) | $J_{sc}$ (mA/cm$^2$) | $J_{sc\text{-}AVE}$ (mA/cm$^2$) | $V_{oc}$ (V) | FF (%) | PCE (%) | $PCE_{AVE}$ (%) | $R_s$ (Ω) | $R_{sh}$ (Ω) |
|---|---|---|---|---|---|---|---|---|
| 2.5×10$^6$ | 20.01 | 19.99±0.02 | 0.95 | 75.4 | 14.33 | 14.14±0.14 | 58 | 25127 |
| 1×10$^6$ | 19.41 | 19.47±0.04 | 0.94 | 74.3 | 13.48 | 13.34±0.12 | 53 | 12445 |
| 0.5×10$^6$ | 19.28 | 19.19±0.10 | 0.91 | 72.1 | 12.69 | 12.80±0.22 | 36 | 8772 |
| 0 | 19.11 | 18.07±1.0 | 0.93 | 72.3 | 12.86 | 12.25±0.51 | 51 | 21920 |
| -0.5×10$^6$ | 18.84 | 17.92±0.65 | 0.93 | 70.5 | 12.28 | 11.71±0.40 | 53 | 19168 |
| -1×10$^6$ | 17.29 | 16.70±0.42 | 0.93 | 67.3 | 10.76 | 10.70±0.06 | 88 | 10167 |
| -2.5×10$^6$ | 16.42 | 16.35±0.22 | 0.92 | 66.8 | 10.14 | 9.57±0.39 | 57 | 9970 |

3.1 Performance of the most PCE cell

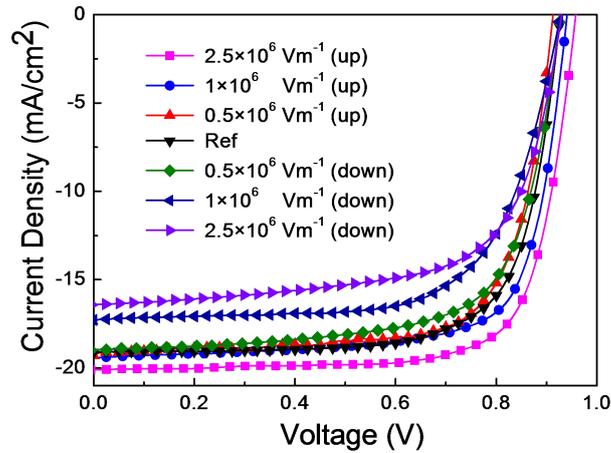

Figure 2. J–V curves of perovskite solar cells with with different EEFs under AM 1.5G illumination of 100 mW cm$^{-2}$.

Figure 2 shows JV curves of the cells whose PCE is the most one in every group. Have a clear vision, the effects of the positive/negative EEFs on JV measurements.

Figure 3(a) shows Voc of the cells whose PCE is the largest one in a group as a function of EEFs. Both of the positive EEFs of 1.0×10$^6$ V/m and 2.5×10$^6$ V/m can produce significant increase in Voc, but the negative EEFs do not change Voc obviously, even the positive EEF of 0.5×10$^6$ V/m degrades Voc. It indicates that it maybe is a accidental in changing Voc by EEF treatment. For strong EEF, the larger Voc may be caused from the larger Jsc and the enlargement of the layer band gap which is caused from the positive strong EEF Voc [21].

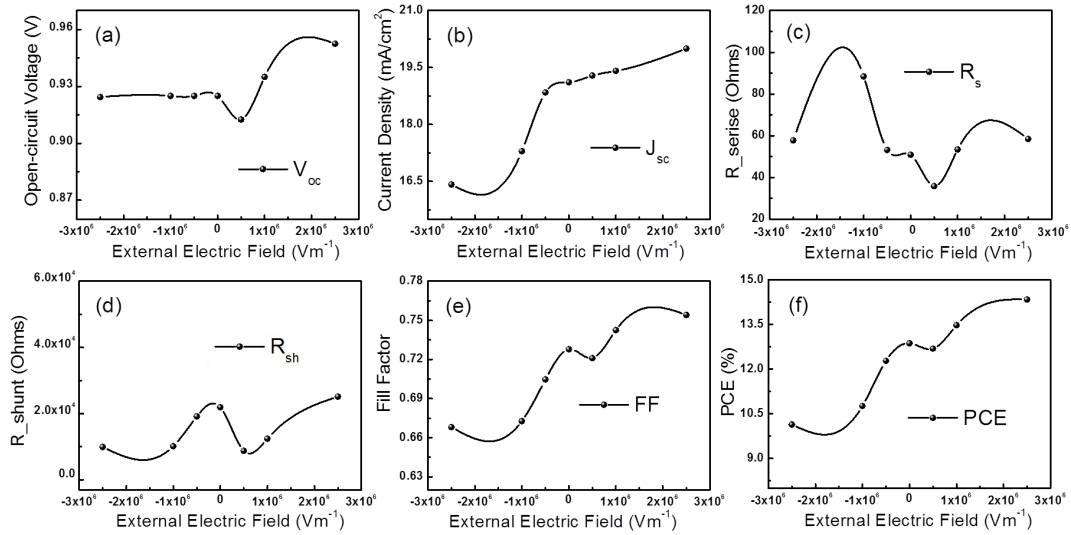

Figure 3. The correlation between the performance parameters, (a), Voc; (b), Jsc; (c), Rs; (d) Rsh; (e), FF; and (f) PCE, as a function of EEF.

Figure 3(b) shows the correlation between Jsc and EEFs. Without exception, Jsc has a nice positive correlation with EEFs, i.e., the positive EEFs increase Jsc, but the negative EEFs reduce Jsc, respectively. The largest positive EEF increases Jsc from 19.11 mA/cm$^2$ to 20.0 mA/cm$^2$, while the largest negative EEF reduces Jsc to 16.42 mA/cm$^2$. Both of the extent of variation reaches 4.7% and -14.9%, respectively.

As a reason of the positive correlation between EEFs and Jsc, it is necessary to analyze the roots. First, With the applied EEF in annealing process, the polar perovskite crystal unit is readily movable and almost certainly to exhibit a polarization, and is arranged in a more uniform ordering heterojunction. Because the applied EEF was kept constant till the annealing temperature drops down to room temperature, the polar crystal units are arranged orderly by EEF to form a more regular crystallization. The nice crystallization can be useful for generation of exciton and carriers transport, which serves as a good basis for photocurrent density.

However, the negative EEF treatment which creates a deterioration in cell performance demands us to find the deep physical mechanism of the applied EEF.

OPIC has been reported to have ferroelectric property with a low coercive field which enables the switching polarization during JV measurement[21]. Another study claims that the ionic migration is confirmed to be the main cause of JV hysteretic behavior occurred in OPIC devices[17]. In our opining, both the ferroelectricity or ionic migration can induce the displacement polarization in the perovskite film, which the conceptual diagram is shown in Fig. 1. The polarization-induced electric field inside OPIC layers should play an important role in changing the physical parameters of the solar cells, including the regulation of the interface band structure of the heterojunction. The positive EEF not only aligns the crystal unit orderly, but also produces a displacement polarization field along with the direction of the original BIF, which increases the net electric field and expands the width of depletion region of the junction. As a result, the enhanced BIF makes the driving force of charge separation and transport be larger, and give an obvious improvement on Jsc.

For negative EEFs, the field brings an opposite effect on BIF, i.e., the width of depletion region of the junction becomes thinner, therefore results in a reducing on the driving force of charge separation and transport. And hence the reducing of Jsc.

In Fig. 3 (c) and (d), the series and shunt resistance (Rs, Rsh) did not emerge obvious correlation with EEFs treatment. The order crystallization of the perovskite layer can not induce an effective and regular reducing or increasing in Rs and Rsh. It indicates the contingency on Rs and Rsh in manufacturing the solar cell. This also illustrated that Rs did not give contribution on the improvement of Jsc by positive EEF.

With FF, Fig. 3 (e) presents a positive correlation as a function of EEFs, except the positive EEF of $0.5 \times 10^6$ V/m. In case of increase of Jsc and irregular Rs and Rsh, the increase of FF with increasing EEFs indicates that the carrier lifetime is improved because of the enhancement of BIF caused by the positive EEFs. For the negative EEFs, they did the opposite effect.

As a composite effect, the increases of Jsc, FF and part Voc caused by the positive EEFs bring an improvement on PCEs, while the negative EEFs result in a deterioration on PCEs which are shown in Fig. 3 (f). One can find that the functions between PCE and EEFs reveal a fine positive correlation. For the largest positive EEF, PCE achieved 14.33, whose improvement reached 11.4% compared with non-EFE sample whose PCE is 12.86; but the largest negative EEF resulted in a degradation with percentage of 21.2% compared with the reference cell. Therefore, the effect of EEF treatment is obvious and interesting.

3.2 The average effect of EEFs on performance improvement

Figure 4 shows the average data with standard deviation of the cell performance as a function of EEFs. For Voc shown in Fig. 4(a), the weak EEF treatment did not bring regular changes compared to the reference cell, but the largest positive and negative EEFs produce the largest and the smallest Voc 0.95V and 0.89 V, respectively. However, the standard deviation of the data for the largest negative EEF is larger, which indicates the contingency on improving Voc.

Figure 4 (b) shows the correlation between Jsc and EEFs. Just like the data of the most large PCE, Jsc changes sensitively with the changing of EEF treatment by a positive correlation. The average extent of variation of Jsc reaches 10.6% and -9.5%, respectively. Except the no-EEF, the less standard deviations of the data indicate that the effect of improving Jsc is stability and reliability.

For Rs and Rsh shown in Fig. 4(c) and (d), they did not turn out obvious correlation with EEF treatment, and the standard deviations are larger. It also indicates that Rs and Rsh are contingency on cell manufacturing.

In Fig. 4 (e) and (f), both the functions of FF and PCE with EEFs reveals the prominent and fine positive correlations with small standard deviations. For the largest positive EEF, PCE of the cell was improved from 12.25 to 14.14, whose improvement equals 15.4% compared to the reference cells. The largest negative EEF only obtained a PCE of 9.57, which changes PCE in percentage of 21.9%. Therefore, the comprehensive effect of EEF treatment on PCE improvement is also stability and reliability, just like Jsc.

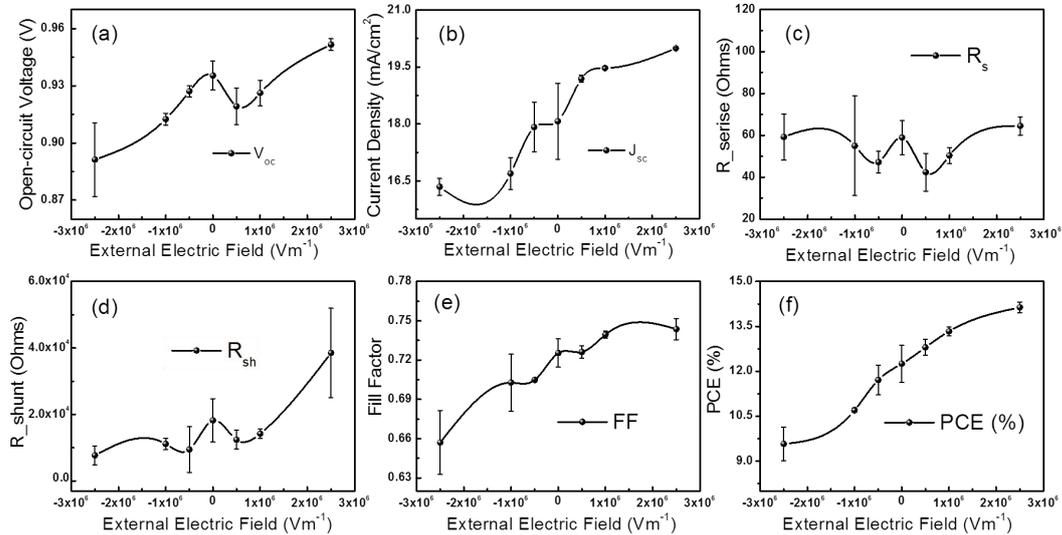

Figure 4. The correlation between the average data of performance parameters, (a), Voc; (b), Jsc; (c), Rs; (d) Rsh; (e), FF; and (f) PCE, as a function of EEF.

4. Conclusion

As a conclusion, an efficient method, which is able to improve the physical properties including FF, Jsc and PCE of the organometal halide perovskite solar cell, is proposed. In the annealing process of the perovskite absorb layer, the polar ions of the perovskite crystal unit are readily movable and certainly exhibits a polarization in external electric field. The ionic migration forms a displacement polarization field which can enhance or weaken the intrinsic BIF existed in the perovskite heterojunction harmonizing the hole and electron transport layers of the solar cell. The enhanced net electric field expands the width of depletion region of the junction to make the driving force of charge separation and transport, even the carrier lifetime being larger. Therefore, it results in an significant and stable improvement on Jsc, FF, and PCE finally, whether for the best cells or for the statistics average data. However, the improvement on Voc is inconclusive. The study proposed a physical process in modifying the efficiency of OPIC devices and provides new evidence about the reason on current-voltage hysteresis of the peroveskite solar cell.

Acknowledges

We acknowledge financial support from the Natural Science Foundation of China (Nos 61307036 and 11074066).